\documentclass[%
reprint, %linenumbers,
groupedaddress,
 amsmath,amssymb,
 aps, physrev,
]{revtex4-2}

\usepackage{graphicx}% Include figure files
\usepackage{dcolumn}% Align table columns on decimal point
\usepackage{bm}% bold math
\usepackage{multirow}%
\usepackage{amsmath,amssymb,amsfonts}%
\usepackage{amsthm}%
\usepackage{mathrsfs}%
\usepackage[title]{appendix}%
\usepackage{xcolor}%
\usepackage{soul}    % 提供 \hl 等高亮命令
\usepackage{textcomp}%
\usepackage{booktabs}%
\usepackage{algorithm}%
\usepackage{algorithmicx}%
\usepackage{algpseudocode}%
\usepackage{listings}%

\usepackage[main=english]{babel}

\begin{document}

\title{Broadband Tunable Photon-Pair Generation and Spectrum Measurement Based on Noncritical Lithium Niobate Crystals}

\author{Zhao-Qi-Zhi Han$^{1,2,3}$}
\altaffiliation{These authors contributed equally to this work.}

\author{Bo-Wen Liu$^{1,2,3}$}
\altaffiliation{These authors contributed equally to this work.}

\author{He Zhang$^{1,2,3}$}
\email{zhanghe23@ustc.edu.cn}

\author{Zhi-You Li$^{1,2,3}$}

\author{Xiao-Hua Wang$^{1,2,3}$}

\author{Jin-Peng Li$^{1,2,3}$}

\author{Zheng-He Zhou$^{1,2,3}$}

\author{Qi-Yu Chen$^{1,2,3}$}

\author{Yin-Hai Li$^{1,2,3,4}$}

\author{Zhi-Yuan Zhou$^{1,2,3,4}$}
\email{zyzhouphy@ustc.edu.cn}

\author{Bao-Sen Shi$^{1,2,3}$}
\email{drshi@ustc.edu.cn}

\affiliation{%
$^{1}$Laboratory of Quantum Information, University of Science and Technology of China, Hefei 230026, China\\
$^{2}$CAS Center for Excellence in Quantum Information and Quantum Physics, University of Science and Technology of China, Hefei 230026, China\\
$^{3}$Anhui Province Key Laboratory of Quantum Network, University of Science and Technology of China, Hefei 230026, China\\
$^{4}$Anhui Kunteng Quantum Technology Co. Ltd., Hefei 231115, China
}

\date{\today}

\begin{abstract}
Photon pairs play a vital role in modern science, driving extensive research into their generation. Yet, the narrow phase-matching bandwidth of conventional crystals has largely confined studies to specific wavelengths, leaving research on broadband tunable sources underexplored. Here, we employ a non-critical phase-matched lithium niobate (LN) crystal to generate widely tunable photon pairs.
The generated near-infrared (NIR) photon pairs exhibit a high coincidence-to-accidental ratio $(CAR >20\ dB)$ and are tunable across the $800–1600 nm$ range. 
We further showcase the utility of NIR photon pairs in spectroscopy by detecting carbon monoxide (CO) gas absorption. This approach will facilitate the design of advanced LN-based photonic experiments.

\end{abstract}

\maketitle

\section{INTRODUCTION}
Photon pairs are indispensable resources in modern photonic quantum technologies. Their nonclassical correlations, manifested in entanglement and strong time–energy/position–momentum correlations, enable a broad set of capabilities ranging from quantum secure communication\cite{Paraiso2021} and quantum computing\cite{DiCarlo2009,Liu2018} to correlation-based sensing\cite{seth2008, Barzanjeh2015,wolley2022} and imaging\cite{Cui2023,Olivieri2020}, also giving rise to some fields such as computational ghost imaging\cite{Shapiro2008} and quantum LiDAR\cite{Pittman1995,Hardy2013}. In parallel, the NIR spectral region technology has become a key working window for both communication and sensing, supporting applications such as fiber-optic communication\cite{MEARS1987}, biomedical analysis\cite{Zhu2024,Vahrmeijer2013,Zhang2024} and autonomous systems \cite{Han2016,Sibi2016}. These trends motivate photon-pair sources that are not only high-quality, but also flexibly addressable across wide wavelength ranges.

To date, correlated photon pairs are predominantly generated via nonlinear optical processes such as spontaneous parametric down-conversion (SPDC) in second-order nonlinear crystals or spontaneous four-wave mixing (SFWM) in third-order platforms\cite{BURNHAM1970,Rarity2005}. While these methods are well established, most existing sources are optimized for narrow phase-matching bandwidths, often operating within fixed wavelength bands. This limitation restricts their utility in applications requiring broadband tunability, such as wavelength-division multiplexing (WDM)\cite{Halder2009}, spectral measurement\cite{JOBSIS1977,Pasquini2018}, and systems that must adapt to varying absorption lines or transmission windows. Furthermore, approaches that pursue very large bandwidths through dispersion engineering \cite{Kues2017,Chen2024} and waveguide designs\cite{Zhao2025} can complicate device design and often yield short correlation times, which is not ideal for time-correlation-based measurements.

A complementary route is to realize frequency-tunable broadband photon pairs by dynamically varying the phase-matching condition, so that wavelength agility is achieved without sacrificing the intrinsic two-photon correlation structure. This approach not only achieves a broad bandwidth but also preserves the frequency correlation between the photons, thereby offering distinct advantages for time-insensitive applications. In this context, noncritical phase matching (NCPM) in LN is particularly attractive: the phase mismatch depends only weakly on wavelength, allowing continuous, wide-range tuning through temperature control. Additionally, NCPM offers practical benefits over critical phase matching, including the avoidance of spatial walk-off\cite{Boyd1968}, which can significantly reduce conversion efficiency, and the mitigation of thermal lensing effects\cite{Weber1999}, leading to more stable and efficient photon-pair generation.

In this work, we present a widely tunable, broadband photon-pair source based on Type-I SPDC in an x-cut LN crystal pumped by a $532\ nm$ continuous-wave (CW) laser. By operating the crystal under NCPM conditions at temperatures up to 250$^\circ C$, the generated photon pairs can be tuned across the NIR spectrum  from $800\ nm$ to $1600\ nm$, while maintaining a high coincidence-to-accidental ratio $(CAR > 20\ dB)$. 
In addition, the measured second-order correlation function exhibits a temporal width approximately $0.5\ ns$, which enables high-precision time calibration.
We further demonstrate how the combination of wide spectral tunability, photon-level detection, and robust temporal correlation can be leveraged for practical measurements, like gas spectroscopy, by detecting CO absorption. These results establish an LN-based, temperature-tunable photon-pair platform that simultaneously offers wide wavelength coverage, high signal-to-noise coincidence performance, and useful time-correlation features, making it promising for broadband spectroscopy and correlation-assisted optical metrology in the NIR.

\section{THEORETICAL METHOD AND EXPERIMENTAL SETUP}
In our type-I NCPM SPDC process, the nonlinear crystal is oriented to be pumped along a specific direction $\phi=0 ^{\circ},\theta=90^{\circ}$ aligned with the optical axes of the crystal. In this case, the nonlinear coefficient $d_{31}$ of LN could be maximally utilized. Based on the energy conservation condition (${\lambda _p} = \frac{{{\lambda _s}{\lambda _i}}}{{{\lambda _s} + {\lambda _i}}}$), the phase mismatch for collinear NCPM SPDC can be expressed as

\begin{equation}
    \Delta k(T)=2\pi\left [\frac{n_p^e(\lambda_p,T)}{\lambda_p}-\frac{n^o_s(\lambda_s,T)}{\lambda_s}-\frac{n^o_i(\lambda_i,T)}{\lambda_i}\right ]
\end{equation}

where, $\lambda_i \ (i=p,s,i)$ are the wavelengths of the pump, signal and idler beams and $n^j\ (j=o,e)$ refer to the refractive indices of ordinary and extraordinary beams, respectively. 
Given that the second-harmonic generation of $1064\ nm$ at $108.9^\circ C$ was observed,  by controlling the temperature above this point, the phase matching of two new wavelengths will be automatically satisfied towards both sides. Consequently, broadband photon pairs spanning $800-1600\ nm$ can be efficiently generated.

For the gas absorption experiment to be demonstrated, numerical analysis can be performed based on the Beer-Lambert law. The intensity of a laser beam after transmission through the absorbing gas medium can be expressed as\cite{Cope1988}:

\begin{equation}
    I=I_0 \mathrm{exp}\left[-\alpha(v)L\right]=I_0 \mathrm{exp}\left[-S(T)\phi_v PX_{abs}L\right]
\end{equation}

where $L$ is the absorption path length, $I_0$ is the light intensity achieved without absorption, also the overall intensity background. $\alpha(v)=S(T)\phi_v P x_{abs}$ denotes the absorption coefficient, where $S(T)$ refers to the absorption line strength, $\phi_v$ is the normalized line-shape function, P is pressure and $x_{abs}$ is the mole fraction of the absorbing gas. This equation describes the absorption intensity of gas for monochromatic light. 
For light with an actual spectral line shape $g(v)$, the transmitted intensity still can be calculated via convolution:
\begin{equation}
    I=I_0\int g(v)\mathrm{exp}[-\alpha(v)L]dv
\end{equation}

Additionally, considering the temporal correlation of photon pairs and its potential role in optical path length detection, when the path length of one photon is fixed, minute variations in the optical path difference $\Delta x$ can be inferred from the time delay $\Delta \tau$ observed in the coincidence measurement of the photon pairs as $\Delta x=c\cdot \Delta\tau$.

\begin{figure*}[t]
    \centering
    \includegraphics[width=0.9\linewidth]{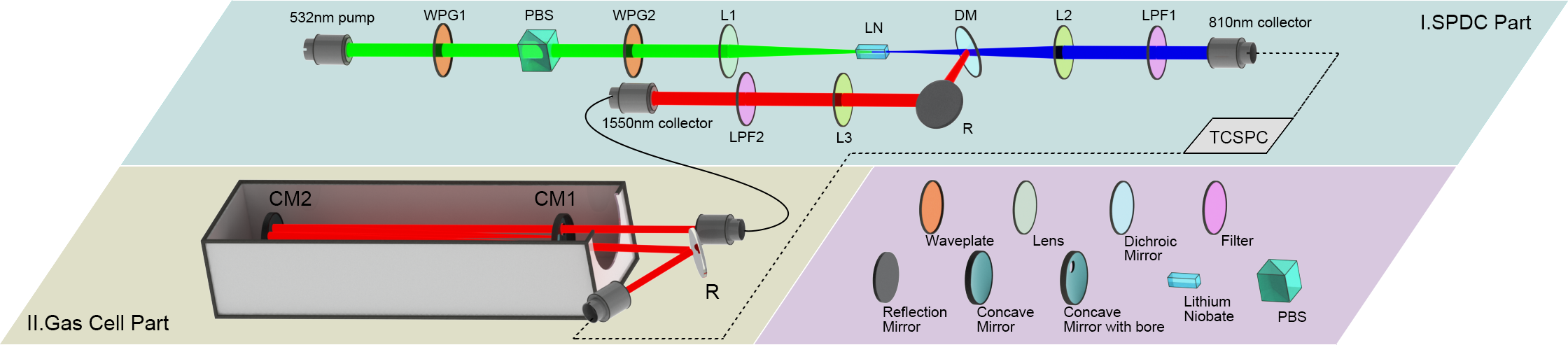}
    \caption{Schematic of the experimental setup. L: lens. WPG: waveplate group, consisting of a half-wave plate (HWP) followed by a quarter-wave plate (QWP). M: mirror. DM: dichroic mirror, with the cut-on wavelength of $1100nm$. LPF: longpass filter. The SPDC Part shows the generation of photon pairs and a temperature control system to monitor the temperature of LN crystal, not shown in the figure. The Gas Cell Part shows the path of the beam, which undergoes multiple reflections within the gas cell before exiting through an aperture in a concave mirror for collection.
    }
    \label{fig1}
\end{figure*}

In this work, a photon-pair source for NIR band based on Type-I SPDC NCPM in a $20\ mm$ x-cut LN crystal was built. The schematic for the experimental setup is shown in \textbf{Figure \ref{fig1}}.
The $532\ nm$ pump light was emitted from a frequency-stabilized, frequency-doubled Nd:YAG continuous-wave (CW) laser and propagated through a $2\ mm$ single mode fiber (SMF) coupler. The waveplate group (WG), which includes a QWP and a HWP, serves as a crucial part in controlling the initialization of the system. The first WG was to control the input power of the system with the following PBS, while the second WG modified the polarization for the SPDC process. In our expected $e\xrightarrow{}o+o$ phase matching, the extraordinary light was the ideal input mode. The following lens with the focal distance $F=100\ mm$ was used to ensure a high power density in the LN crystal. The beam waist in LN crystal was approximately $w=150\ \mu m$ in the center, which ensured good quality of beams for high efficiency in frequency conversion. The LN crystal was placed in a temperature-controlled system, which could retain the desired temperature with fluctuations below $0.01\ K$, as detailed in \textbf{Supplementary Information II}. The photon pairs with non-degenerate wavelengths were generated through Type-I SPDC process. We refer to the longer-wavelength photons as near-infrared (NIR) photons and the shorter-wavelength ones as visible (VIS) photons. After being separated into two parts by a $1100\ nm$ cut-on longpass dichroic mirror, photon pairs were refocused with respective lenses, which actually formed two $4-f$ optical systems, and were then coupled into each fiber. 

The photons with longer wavelength were coupled into a sealed cavity and reflected between two concave mirrors multiple times. A calcium fluoride ($CaF_2$) lens (not shown in the schematic), which has excellent transmittance in NIR band\cite{Daimon2002}, was used as the input window in the gas cell. A concave mirror with a central aperture (diameter $\Phi=30\ mm$) allowed the incident beam to enter and the attenuated beam to exit the multipass cavity. Subsequently, the VIS and NIR photons were then detected by separate single-photon counting modules (SPCM, SPCM-AQRH-14 manufactured by Excelitas). The photon detection efficiency (PDE) exceeds $60\%$ in the wavelength of $800\ nm$ and remains at approximately $10\%$ near the edge band around $1000\ nm$.
The PDE of SPCM for NIR photons (ID220-FR from ID Quantique company) remained above $15\%$ within the $1200-1500\ nm$ band and plunged sharply around $1600\ nm$.
The signals were processed by a time-correlated single-photon counting (TCSPC) system for coincidence measurements, which outputs data through different channels. The output data $N_{long}, N_{short}$ respectively refer to NIR and VIS non-degenerate photon pairs, and $N_{cc}$ is the coincidence count. 

During long-term measurements over several hours, power fluctuations are unavoidable due to various factors. For example, the zero drift induced by environmental temperature changing over time, and the sudden increased noise caused by the switching power supply. They significantly impact performance in practical applications but combining measurements with data from other channels will significantly improve the performance. The absorption signal for gas in the $N_{long},N_{cc}$ and can be normalized using $N_{short}$ to eliminate the influence of input power, as in $N_{nor}=N_{cc}/N_{short}$.

Extensive research has demonstrated that many types of gases have absorption spectra in NIR band\cite{Thompson1946}, and various technologies have been developed for gas detection in these bands\cite{Wong2011}. Among studies related to common chemical gas, CO occupies a significant position due to its role in organic chemistry, industry and its associated hazards. In many instances, CO is primarily considered as a noxious gas due to its adverse effect on both organisms and the environment. Notably, there exhibits an absorption peak between $1.558\ \mu m$ and $1.613\ \mu m$ for CO according to the HITRAN database\cite{Hitran2022}, which falls within the tunable band in this paper. In this case, CO was chosen as the absorbing material with more reasons.
Furthermore, the helium (He) as an inert gas possesses minimal absorption peaks, which mainly occur in extreme ultraviolet (EUV) and VIS band. The spectral lines of He I were obtained from the NIST Atomic Spectra Database \cite{NIST_ASD}. The only absorption peak for He in NIR was near $1083nm$, which does not affect the absorption of CO. Thus, He was chosen as standard gas to serve as the benchmark.

It should be noted that, based on the vibrational-rotational spectrum of CO, the absorption peak bandwidth will be extremely narrow, on the order of $10^{-2}\ nm$. And for SPDC process using a $10\ mm$ nonlinear crystal, the generated photon bandwidth will be near $2\ nm$, which means its bandwidth would cover multiple absorption peaks, not to mention the measurement interval of temperature is limited. Hence, the discrete details of absorption lines will be difficult to resolve directly. For concreteness, the entire contour of P- and R-branches could be roughly observed while the discrete peaks are overwhelmed by broadband absorption. Nevertheless, mathematical tools to extract as much detail as possible still exist. We have mentioned an efficient algorithm for extracting discrete energy levels in the theoretical method section.

\section{EXPERIMENTAL RESULTS}
\begin{figure*}
    \centering
    \includegraphics[width=0.7\linewidth]{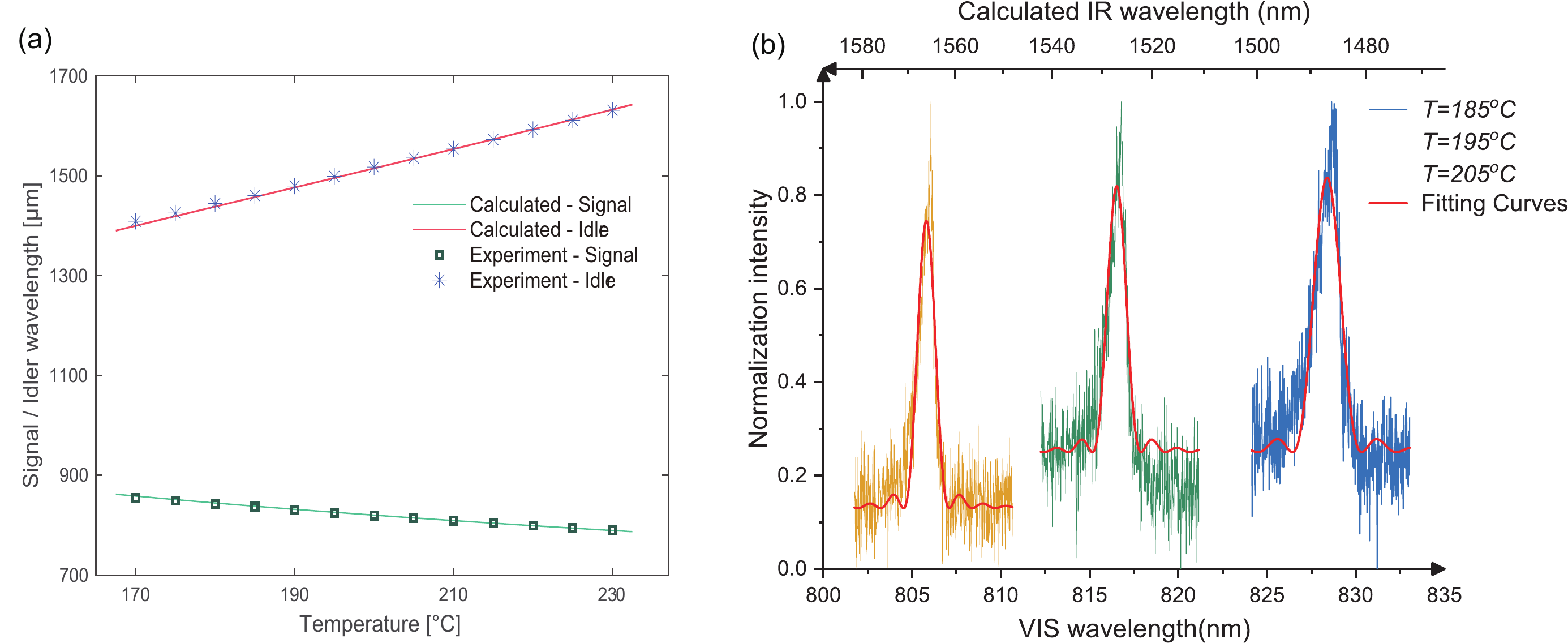}
    \caption{The schematic representation of the photon-pair characterization setup. (a) Dependence of the center wavelength of the photon pairs on the LN temperature. (b) The spectral bandwidth distribution of VIS photon pairs generated at different temperatures. The upper axis represents the wavelength of NIR photons calculated through energy conservation.}
    \label{fig2}
\end{figure*}

\begin{figure*}
    \centering
    \includegraphics[width=0.7\linewidth]{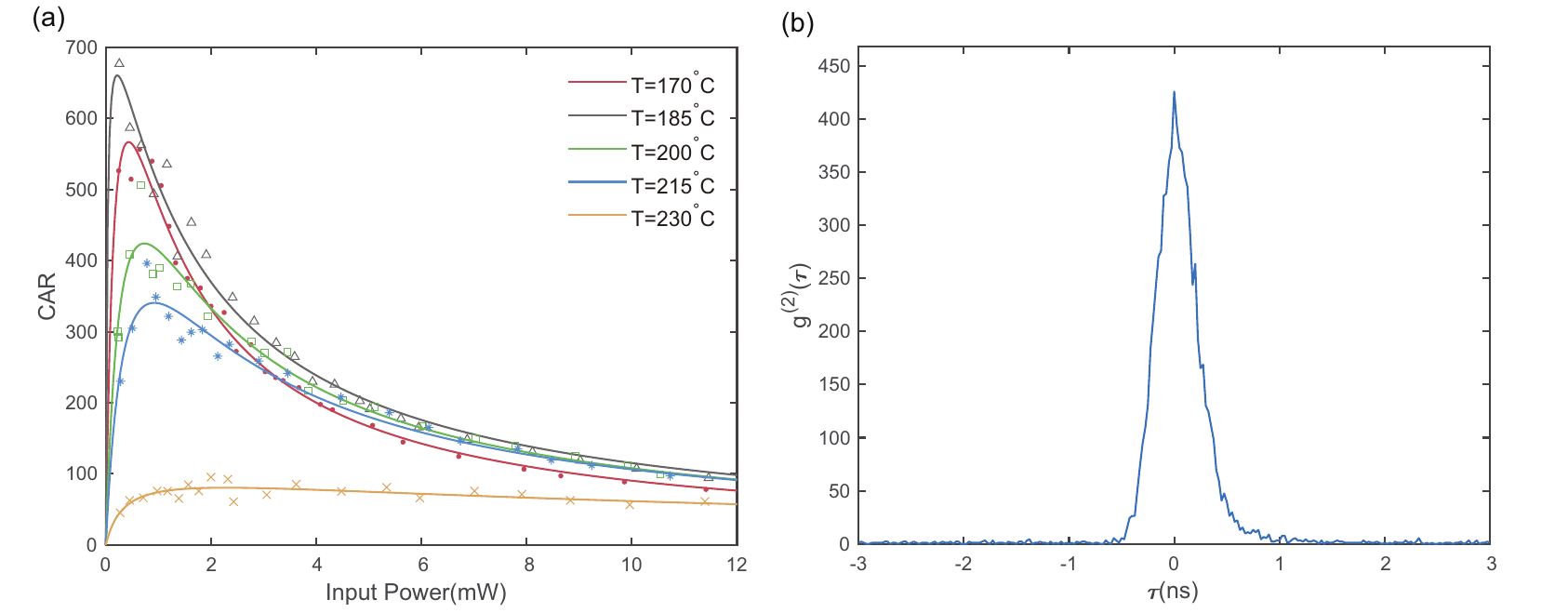}
    \caption{(a) The CAR curves obtained from several temperatures at equal intervals. The horizontal axis represents the power input to the crystal, and the vertical axis CAR represents the signal-to-noise ratio of photon pairs. (b) The pattern obtained by normalizing the coincidence curve collected from TCSPC. For ease of observation, the peak position is set as $\tau=0$ on the horizontal axis.}
    \label{fig3}
\end{figure*}

\textbf{Photon pairs quality.}
The photon pairs for $800-1600\ nm$ were generated using a Type-I noncritical SPDC process and the characterization of this photon-pair system was shown in \textbf{Figure \ref{fig3}}. 
The temperature-wavelength tuning curve plotted in \textbf{Figure \ref{fig2}(a)} was calculated based on the previous research \cite{Liu2025} and was experimentally verified in the wavelength band where a large separation exists. Over the temperature range of $170^\circ C$ to $230^\circ C$, the maximum wavelength offset between theory and experiment was observed to be below $3.4\ nm$ for signal photons, which falls within acceptable error margins. At low temperature, the experimental results would be more consistent with the calculated values, as the small change for refractive index introduced less error. We collected the spectrum of VIS photons to analyze their bandwidth and representative results were presented in \textbf{Figure \ref{fig2}(b)}. To clearly illustrate the distribution of photon bandwidth, we normalized the collected photon spectrum signals. The coordinate axis below, labeled 'VIS wavelength', corresponds to the wavelength of VIS photons and the upper axis represents the wavelength of NIR bands calculated from energy conservation. For the collected VIS photons shown in this figure, the bandwidths were distributed within $1.6-2.1\ nm$. Although the theoretically calculated bandwidth is around $1.2\ nm$, the focusing of the laser beam resulted in the reduction of the effective interaction length in LN crystal. The effective length was much shorter than the physical $2\ cm$ length, resulting in an actual VIS photon-pair bandwidth of approximately $1.6\ nm$. Additionally, the spectral broadening observed in the figure was further increased by the  $0.1\ nm$ sampling resolution of spectrometer. When converted to the bandwidth of VIS photons to the NIR ones, this became approximately $4\ nm$.

To characterize the signal-to-noise ratio of the photon-pair source, the Coincidence-to-Accidental Ratio (CAR) curves for VIS photons measured with a time resolution of $0.025\ ns$ were plotted in \textbf{Figure \ref{fig3}(a)}. The CAR is the ratio of coincidence counts to dark counts and represents a direct measure of both single-photon purity and the system signal-to-noise ratio.
Clearly, the temperature significantly affected the signal-to-noise ratio, with the CAR dropping below $20\ dB$ when the maximum temperature was set to $230^\circ C$. 
The primary reason for this phenomenon was that the corresponding wavelength had reached the cutoff edge of the single-photon avalanche diode (SPAD). The rapid decline in detection efficiency led to the sharp reduction in the CAR.
To achieve a high CAR and maximize the output power, thereby enabling the observation of sufficiently pronounced phenomena in gas absorption, we conducted tests at an input power of approximately $1\ \mathrm{mW}$.
Although there were some fluctuations, the CAR was acceptable in most cases when a reasonable accumulation time was chosen. 

The second-order correlation function $g^{(2)}(\tau)$ of generated photon pairs was shown in \textbf{Figure \ref{fig3}(b)}. We calculated the statistical averaging $C_{bg}$ away from the coincidence position to normalize the original coincidence curve $C(\tau)$ as $g^{(2)}(\tau)=\frac{C(\tau)}{C_{bg}}$. The $0.5\ ns$ bandwidth of the measured correlation function enabled accurate time and position calibration, though ultimate precision was limited by detector time jitter.

\textbf{CO gas absorption.}
The desired pure gas was filled into the evacuated gas cell, with the internal pressure subsequently maintained at approximately one atmosphere. When it was necessary to replace the gas, the cell was purged multiple times to ensure the required purity. The temperature was set successively in the interval of $0.2\ K$, ensuring sufficient sampling density. In order to observe the pronounced absorption, we adjusted the position and orientation of two opposing  concave mirrors in gas cell, until more than 10 light spots appeared on a single concave mirror, which meant the light passed through the gas cell over a path length more than 20 times the length. In fact, an additional $4-f$ optical system was set before the $1550nm$ collection coupler in order to correct the diffraction caused by the light beam propagating within the gas cell. 

\begin{figure*}
    \centering
    \includegraphics[width=0.7\linewidth]{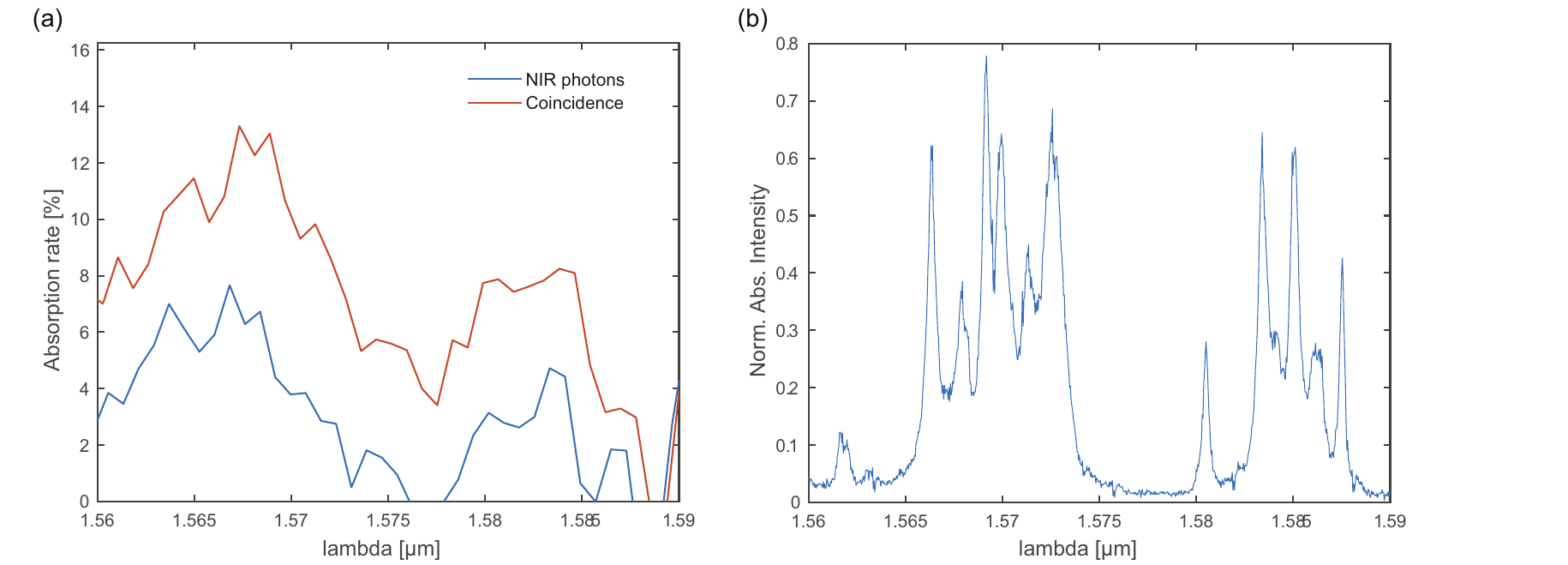}
    \caption{(a) The original CO absorption rates for coincidence measurement and single NIR channel measurement. The wavelength range of $1.56-1.59\mu m$, which shows the most obvious CO absorption, is selected. (b) The CO absorption curve after deep learning post-processing, the vertical label \textbf{"Norm. Abs. Intensity"} refers to the normalized absorption intensity.}
    \label{fig4}
\end{figure*}

The original CO absorption rates for coincidence measurement and single NIR channel measurement are shown in \textbf{Figure \ref{fig4}(a)}. The horizontal axis represents the NIR photon wavelength converted from measured temperature and the vertical axis reflects the absorption rates of the collected photons. Given that the wavelength serves as a critical parameter, and to establish a more accurate conversion, we performed additional spectral calibration using the temperature-wavelength tuning curve in \textbf{Figure \ref{fig2}(a)}. In this figure, there is a significant contrast between the two measurement methods in the absorption band of the target. The absorption signal obtained through coincidence channel is much stronger than that obtained through direct measurement, as the NIR signal background caused by background noise is relatively large, which to some extent reduced the absorption rate. Furthermore, since the measurement needed to be compared with the calibration gas curve, it is crucial to eliminate errors caused by laser power fluctuations between the two measurements. After performing power calibration using the directly collected visible photons from one path, coincidence measurement could avoid the effects of power fluctuations during the measurement process, offering better long-term stability. Additionally, according to the numerical calculation based on the theoretical method section, the peak absorption rate reaches $10\%$ around the wavelength of $1.567\mu m$, which is in good agreement with the coincidence measurement results.

To extract absorption details from the data as thoroughly as possible, we employed a convolution neural network to process the obtained results as described in \textbf{Supplementary I}. After processing the target band of $1.56-1.59\ \mu m$, the separated details were further extracted, as illustrated in \textbf{Figure \ref{fig4}(b)}. By analyzing the absorption response of narrow-linewidth peaks to broadband photons, we revealed spectral details beyond the photon bandwidth limit.

\textbf{Optical path measurement.}
The temporal correlation of photon pairs implies the two-photon phase difference, which can be converted to optical path through simple computation, is measurable in a stable system. The measurement precision is predominantly governed by the temporal coincidence window and the spectral correlation bandwidth of photon pairs.

\begin{figure}
    \centering
    \includegraphics[width=0.95\linewidth]{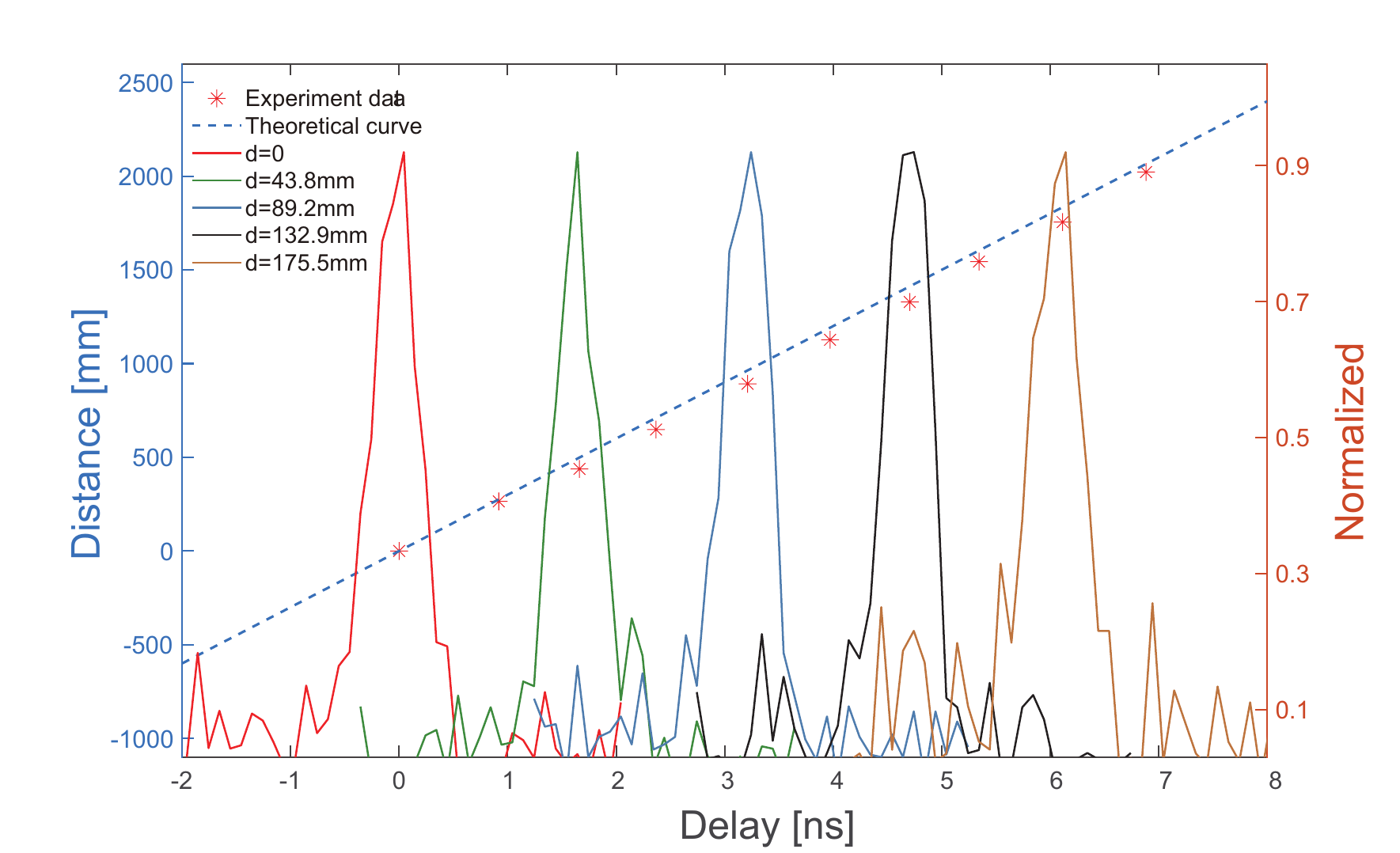}
    \caption{The relation between optical delay and the induced time-delay shifts. The curves were directly obtained from TCSPC at optical delays of $0$, $43.8\ mm$, $89.2\ mm$, $132.9\ mm$, $175.5\ mm$, respectively, and normalized on the right axis for clarity. The dashed line indicates the linear relation between optical delay and time-delay.}
    \label{fig5}
\end{figure}

By controlling the number of photon reflections within the gas cell, the coincidence counts of photon pairs were collected. The prominent change manifested in the time delay of photon coincidences, which reflected the temporal offset between the arrival of the two-photon states at detectors. The correlation between the controlled optical delay changes and time-delay shifts is presented in \textbf{Figure \ref{fig5}}.  Based on the simple relationship of $\Delta x=c\cdot\Delta t$, a dashed line was plotted in figure to verify the function of the optical path detection. The goodness-of-fit between experimental results and theoretical curve yielded a value of $R^2=0.9984$. 

\section{CONCLUSION}
This work provides details of a broadband tunable SPDC photon-pair source for $800-1600\ nm$ based on the noncritical LN crystal. We carried out an overview of the application in the related NIR band which covers almost the entire generated wavelength range. Inspired by the distinctive absorption lines of many gas molecules in NIR band, we decided to use CO gas for optical absorption measurements after thorough investigation. Additionally, considering the correlation features of photon pairs, a measurement scheme utilizing coincidence timing analysis to detect path-length variations was proposed.

Experimentally, the tuning curve and spectral bandwidth of this photon-pair system were tested and the results met expectations. The CAR curves were calculated and the signal-to-noise ratio was above $20\ dB$ in most cases. The prepared longer-wavelength photons were emitted into the gas cell via optical fiber. After multiple reflections within the cell filled with pure CO, the transmitted photons were collected and compared with reference measurements under non-absorbing conditions. This revealed the presence of absorption within the $1.56-1.59\ \mu m$ spectral band. However, due to the broad bandwidth of the photons relative to the absorption lines, discrete absorption peaks could not be directly observed. To extract details of the absorption spectrum as much as possible, algorithmic methods were employed as the analytical approach which is consistent with the physical processes. Post-processing of the data revealed the P and R branches of CO molecular absorption. Furthermore, the measurement of optical path difference yielded the expected results. The analysis presented a well-defined linear relationship between the coincidence time and the optical path difference. The bandwidth of the generated photon pairs can be controlled by adjusting the effective length of the crystal, thereby enabling applications for distance calibration in different scenarios.

We expect that the phase-matching method demonstrated in this work could find utility in other applications, such as driving energy-level transitions in atoms or molecules to serve as quantum interfaces, and extending the system's functionality into imaging for biological studies.

\section{ACKNOWLEDGMENTS}
Zhao-Qi-Zhi Han and Bo-Wen Liu contribute equally to this work. We would like to acknowledge the support from National Key Research and Development Program of China (2022YFB3903102, 2022YFB3607700), National Natural Science Foundation of China (NSFC)(62435018), Innovation Program for Quantum Science and Technology (2021ZD0301100), USTC Research Funds of the Double First-Class Initiative (YD2030002023), Quantum Science and Technology-National Science and Technology Major Project (2024ZD0300800), and Research Cooperation Fund of SAST, CASC (SAST2022-075).

\bibliographystyle{IEEEtran}
\bibliography{reference}

\end{document}

\title{Supplementary Information On Broadband Tunable Photon-Pair Generation and Spectrum Measurement Based on Noncritical Lithium Niobate Crystals}

\author{Zhao-Qi-Zhi Han$^{1,2,3}$}
\altaffiliation{These authors contributed equally to this work.}

\author{Bo-Wen Liu$^{1,2,3}$}
\altaffiliation{These authors contributed equally to this work.}

\author{He Zhang$^{1,2,3}$}
\email{zhanghe23@ustc.edu.cn}

\author{Zhi-You Li$^{1,2,3}$}

\author{Xiao-Hua Wang$^{1,2,3}$}

\author{Jin-Peng Li$^{1,2,3}$}

\author{Zheng-He Zhou$^{1,2,3}$}

\author{Qi-Yu Chen$^{1,2,3}$}

\author{Yin-Hai Li$^{1,2,3,4}$}

\author{Zhi-Yuan Zhou$^{1,2,3,4}$}
\email{zyzhouphy@ustc.edu.cn}

\author{Bao-Sen Shi$^{1,2,3}$}
\email{drshi@ustc.edu.cn}

\affiliation{%
$^{1}$Laboratory of Quantum Information, University of Science and Technology of China, Hefei 230026, China\\
$^{2}$CAS Center for Excellence in Quantum Information and Quantum Physics, University of Science and Technology of China, Hefei 230026, China\\
$^{3}$Anhui Province Key Laboratory of Quantum Network, University of Science and Technology of China, Hefei 230026, China\\
$^{4}$Anhui Kunteng Quantum Technology Co. Ltd., Hefei 231115, China
}

\date{\today}
\maketitle
\clearpage
\tableofcontents
\clearpage

\section{Deep Learning Algorithm Architecture}
The photon pairs generated via the NCPM SPDC process exhibit excellent tunability but often come with a broad photon bandwidth that far exceeds the width of the absorption line. To achieve better measurement performance, a super-resolution model combined with algorithms processing has been demonstrated. We express the measured spectrum in a discrete form:

\begin{equation}
    \boldsymbol{y=D(x\otimes k)+n }
\end{equation}

where, y denotes the low-resolution measurement data, $x$ is the original absorption spectrum, $k$ represents the convolution kernel. The convolution of $x$ and $k$ reflects the physical process of photon absorption and signal broadening. The function $D$ is the models of the entire sampling process and $n$ is the mixture of mixed noise in measurement. The solving model is demonstrated in the following form:

\begin{equation}
\left\{
    \begin{array}{cc}
         &  \mathop{\mathbf{\min}}\limits_{\Theta}\sum_{i=1}^N \mathcal{L}(\boldsymbol{\hat{x}_i},\boldsymbol{x_i})  \\
         &  s.t. \ \ \boldsymbol{\hat{x}_i}= \mathop{\arg\min}\limits_{\boldsymbol{x}} \frac{1}{2\sigma^2}\left |\left |\boldsymbol{y_i-D(x\otimes k)}\right |\right|^2 
    \end{array}
    \right.
\end{equation}

where, $\boldsymbol{x_i}$ is the ground truth label, $\boldsymbol{\hat{x}_i}$ is the predicted value, $\mathcal{L}$ is the loss function and $\Theta$ denotes the trainable parameters of this network. 

By employing a large set of randomly generated spectral curves along with experimentally measured VIS photon waveforms at specific temperatures, the physical measurement process of gas absorption and the superposition of noise within the gas cell can be simulated.

\begin{figure*}[h]
    \centering
    \includegraphics[width=0.8\linewidth]{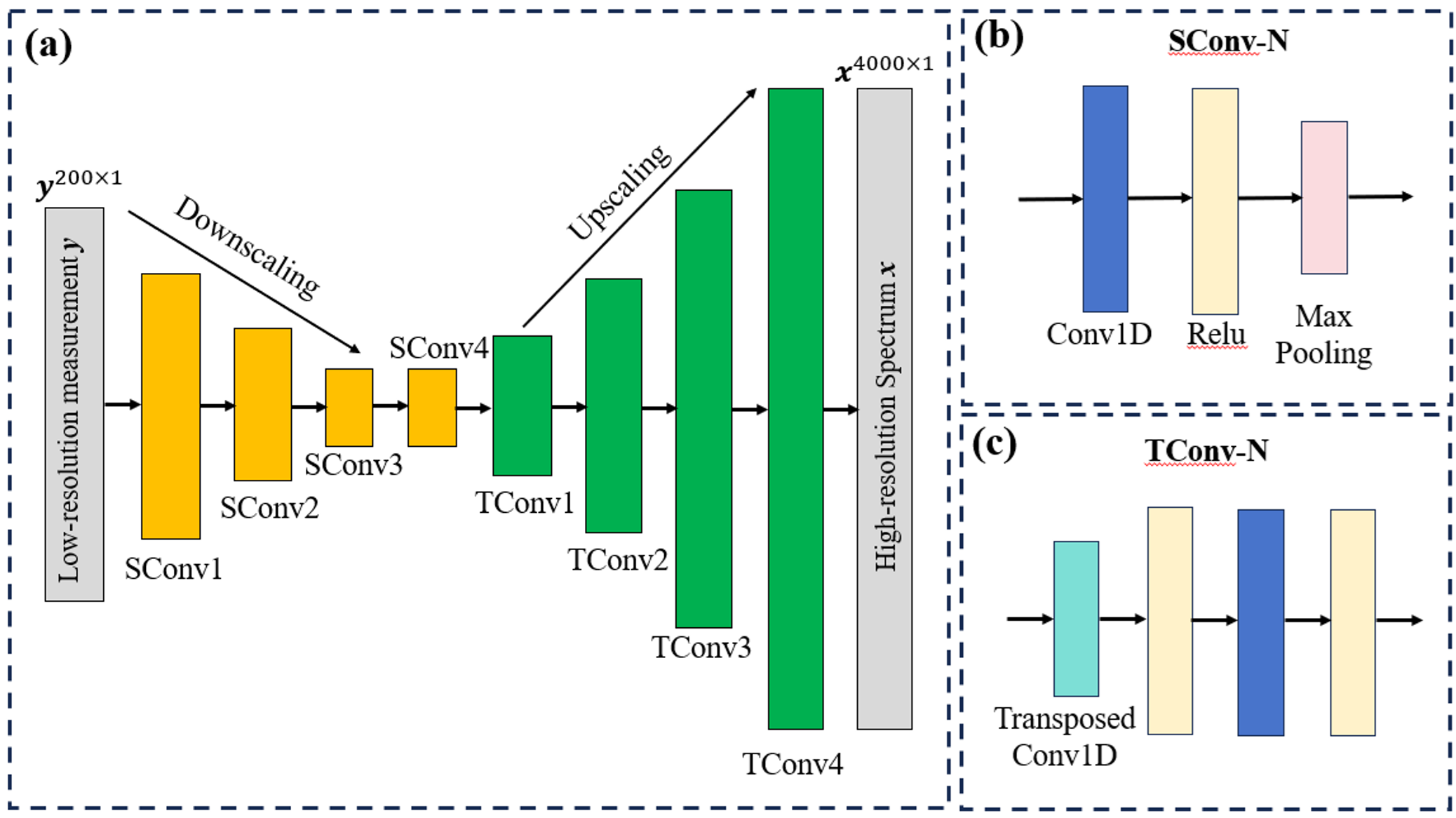}
    \caption{(a) The schematic Diagram of a 1D-CNN Encoder-Decoder Architecture. (b) The schematic diagram of strided convolution Layer. SConV, the strided convolution. (c) The schematic diagram of transposed convolution Layer. TConV, transposed convolution.}
    \label{fig1}
\end{figure*}

This model implements an end-to-end 1D-CNN-based encoder-decoder architecture for spectral data super-resolution and deconvolution, see \textbf{Figure \ref{fig1}}. Assuming that the input low-resolution one-dimensional measured spectrum $\mathbf{y}$ has a data length of $200\times1$, it passes through three downsampling blocks, each typically consisting of convolution, activation, and pooling operations, and one convolutional layer without pooling (SConv4) to expand the receptive field, ultimately resulting in an intermediate layer structure with features of size $25\times1\times512$. Subsequently, the feature map passes through four transposed convolution layers to progressively increase the resolution of the decoder. Each layer includes a corresponding ReLU activation and a standard convolution operation. 

In addition to constructing an effective convolutional network architecture, the preparation of the dataset for training the absorption spectrum deconvolution model must also be addressed. Given the experimental design requirements of this study, it is highly challenging to create a correspondingly large dataset of empirical absorption spectra using multiple known gases covering the mid-infrared range. Accordingly, a large number of randomly generated spectral curves, together with experimentally measured visible photon measurement spectral waveforms obtained at 9 temperature points $T$ (ranging from $185^\circ C$ to $225^\circ C$ at $5^\circ C$ intervals), were used to simulate the physical measurement process of gas absorption of infrared light and noise superposition within an actual gas cell. Directly utilizing the waveforms of photon pairs experimentally generated via the SPDC process (see \textbf{Main Text, Figure 2}), will enable the network to learn the characteristics of the actual measurement system more realistically. The low-resolution result y-dataset and the known ground-truth high-resolution spectrum x-dataset together constitute a standard paired dataset. Within this lightweight data-driven framework, we designed and generated 4000 simulated samples, which is sufficient for a model of this scale.

\section{Temperature Control System Stability}
\begin{figure}[h]
    \centering
    \includegraphics[width=0.5\linewidth]{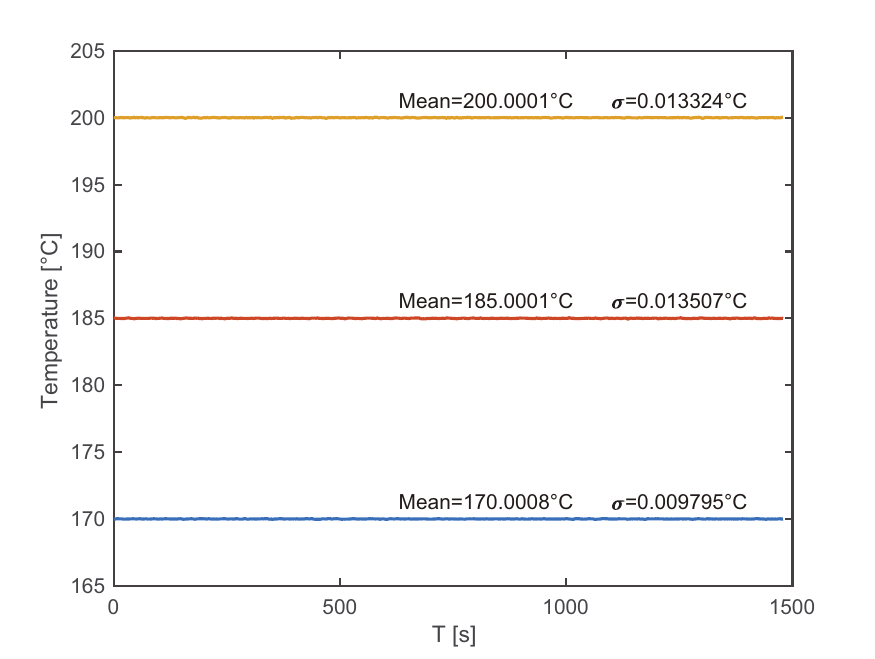}
    \caption{Temperature stability test under prolonged operation. The data for the three temperatures were collected within the same experimental run and were aligned to $t=0$ for clarity.}
    \label{fig2}
\end{figure}

To ensure stable temperature during the SPDC process, we set a sufficiently long waiting time for the automatic temperature switching. In addition, to verify the stability under long-term data collection, we conducted additional  temperature stability tests, as shown in the \textbf{Figure \ref{fig2}}. 

Temperature fluctuations remain around the expected value of 0.01 K. Based on the power output relationship described by:

\begin{equation}
    P_{out}(T)\propto\mathrm{sinc}^2[\Delta k(T)L/2]
\end{equation}

where, $\Delta k(T)=k_p(T)-k_s(T)-k_i(T)$ is the phase mismatch and $L$ is the effective crystal length. Considering the temperature fluctuations, the output fluctuation was maintained below \text{\textperthousand 0.5}. Since the CAR of photon pairs has exceeded 20 dB, these power fluctuations are negligible, and temperature stability can be ensured.

\section{Broadband Absorption}
To better demonstrate the broadband tunability of the photon-pair system, similar absorption measurements were also conducted within the absorption band of water. Given that water absorption is so strong that even the ambient water vapor in air yields a detectable signal, we vented the gas cell to the laboratory atmosphere, thus allowing it to fill with air at the ambient humidity. The measurement band we selected was approximately $1.4-1.52\ \mu m$, lying within the overtone absorption band of water’s vibrational energy levels in the IR region.

\begin{figure}[h]
    \centering
    \includegraphics[width=0.5\linewidth]{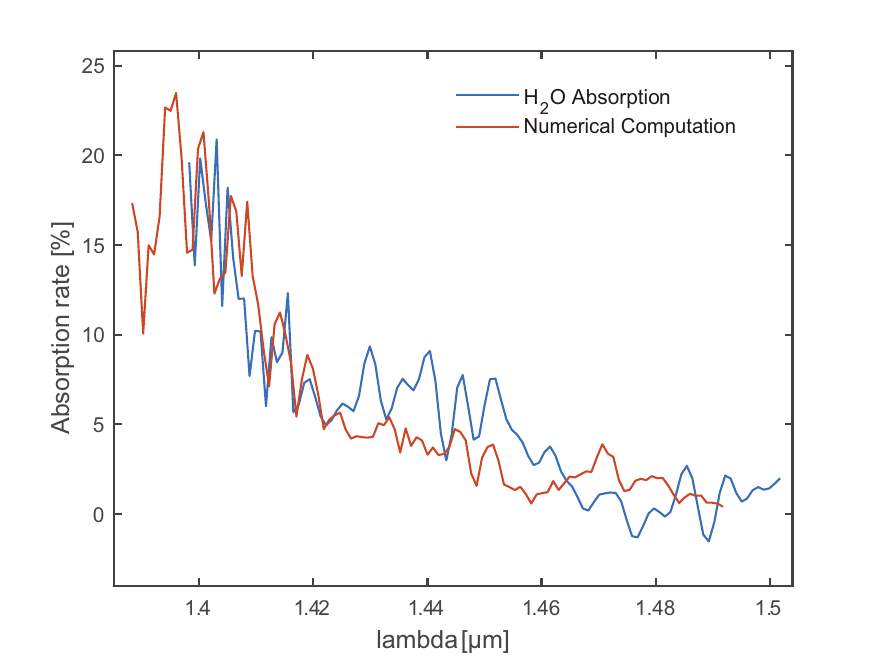}
    \caption{Water absorption results, where the blue line represents the denoised experimental data and the red line indicates the calculated theoretical absorption.}
    \label{fig3}
\end{figure}

This experiment was conducted at an air humidity of $37\%$. With the light beam undergoing 11 reflections within the gas cell, an absorption path length of approximately $4\ m$ was ultimately achieved. The results are shown in the  \textbf{Figure \ref{fig3}}. 
In addition to applying Gaussian denoising to the acquired signal, a theoretical absorption profile was calculated using data from HITRAN with a photon bandwidth of $2\ nm$. Moreover, a minor absorption peak near $1.47\ \mu m$ was not observed in the measured data. The reason for this may be that its absorption intensity was too low, causing the signal to be dominated by the noise.

These data were actually part of a CO absorption measurement. While comparing the absorption data of He, CO, and air in the $1.4-1.6\ \mu m$ region, we identified absorption features attributable to water around $1400\ nm$. Consequently, this spectral segment was not centered on the water absorption peak near $1.37\ \mu m$, but began at approximately $1.4\ \mu m$.

\bibliographystyle{IEEEtran}
\bibliography{supplementary}